\documentclass[12pt,english]{scrartcl}

\usepackage[OT1]{fontenc}
\usepackage[latin1]{inputenc}
\usepackage{amsmath}
\usepackage{graphicx}
\usepackage{babel}
\usepackage{amssymb}
\IfFileExists{url.sty}{\usepackage{url}}
                      {\newcommand{\url}{\texttt}}

\makeatletter

\newif \ifpdf
\ifx \pdfoutput \undefined
      \pdffalse
   \else
      \pdftrue
\fi
\ifpdf
\usepackage{mathptmx}
\usepackage{helvet}
\usepackage{ae,aecompl}
\usepackage[pdftex,urlcolor=blue,colorlinks=true]{hyperref}
\else
\usepackage[ps2pdf,urlcolor=blue,colorlinks=true]{hyperref}
\fi
\begin{document}

\title{Surface plasmon polariton propagation around bends at a metal-dielectric interface}
\author{Keisuke Hasegawa, Jens U. N\"ockel and \href{mailto:miriamd@uoregon.edu}{Miriam Deutsch}\\
\medskip
Oregon Center for Optics,\\
1274 University of Oregon,\\ 
Eugene, OR 97403-1274\\
\url{http://oco.uoregon.edu/}
}

\date{Published in Appl.~Phys.~Lett.~{\bf 84}, 1835 (2004)}

\maketitle
\begin{abstract}
We analyze theoretically the propagation of surface plasmon polaritons about a metallic corner with a finite bend
radius, using a one-dimensional model analogous to the scattering from a finite-depth potential well. We obtain
expressions for the energy reflection and transmission coefficients in the short wavelength limit, as well as an
upper bound for the transmittance. In certain cases we find that propagation on non-planar interfaces may result
in lower losses than on flat surfaces, contrary to expectation. In addition, we also find that the maximum
transmittance depends non-monotonously on the bend radius, allowing increased transmission with decreasing
radius. 
\end{abstract}

Structured materials which allow nanoscale control of light are necessary for achieving compact integrated
photonic devices. While the size of standard optical components and beams is typically set by the diffraction
limit, low dimensional excitations such as surface-plasmon polaritons may be confined to dimensions much smaller
than the wavelength of light. Surface-plasmon polaritons (SPPs), coupled modes of plasmons and photons, are
excited when visible electromagnetic (EM) radiation couples into surface guided modes at metal-dielectric
interfaces \cite{Agranovich,Sernelius}. When propagating along flat interfaces, these are essentially
two-dimensional (2D) waves, with an EM field intensity which peaks at the interface and decays exponentially
into the two adjoining media.

Recently, SPP waveguiding and bending in nano-patterned metallic films were studied \cite{Bozhevlonyi}.
Alternately, it was shown that EM energy may be efficiently transported by near field coupling in plasmon
waveguides comprised of ordered arrays of metal nanoparticles \cite{Aussenegg98}. Optical elements such as
linear waveguides \cite{Atwater}, mirrors, beamsplitters and interferometers \cite{Aussenegg02} were recently
demonstrated.

Interestingly, while significant progress has been made in understanding SPP propagation in nano-structures,
certain fundamental issues pertaining to their guiding on smooth metallic films remain unknown. In particular,
quantifying guiding and energy losses in SPPs propagating around bends in metal-dielectric interfaces is of
great importance, as it should set a limit on feature size in certain plasmonic-circuit devices. Previously, the
problems of refraction \cite{Stegeman} and reflection of SPPs \cite{WallisDawson} at interfaces have been
addressed in this context. In this Letter we present a study of the efficiency of SPP propagation at a curved
metal-dielectric interface in the short wavelength limit.

\begin{figure}[!thb]
\centering{\includegraphics[width=14cm]{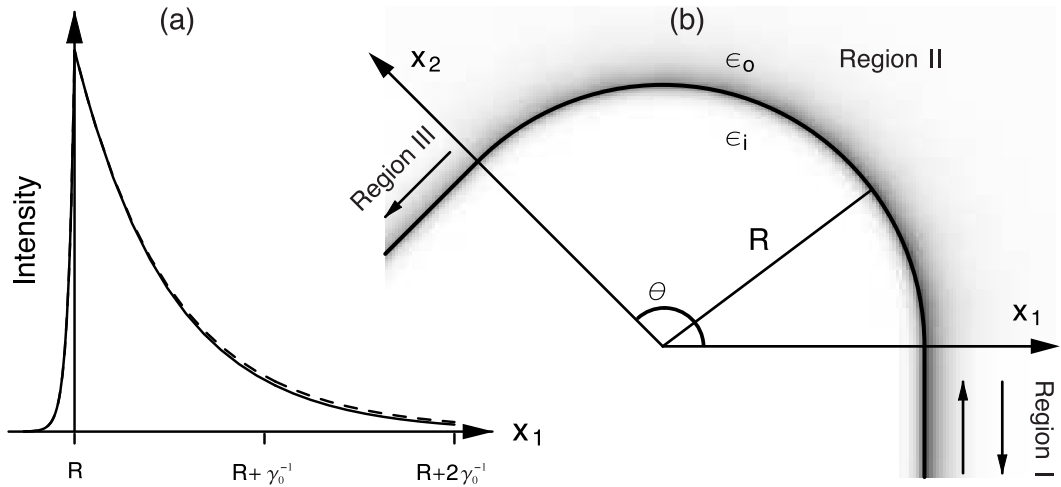}}
\caption{\label{fig1} 
Cross-sectional (a) and top view (b) of the SPP intensity;
  (b) also shows a drawing of the geometry. A metallic corner characterized by a dielectric constant
  $\epsilon_i$ and $\hbox{Re}[\epsilon_i]<0$ has a bend angle $\theta>90^{\circ}$ and a finite bend radius $R$.
  The bend is confined to the region of space shown, with the center of curvature at the origin.
  The rest of space is occupied by a dielectric with $\epsilon_o$. Axes $x_1$ and $x_2$ extend along the boundaries
  between Regions I and II and Regions II and III, respectively. In the $x_1$-$x_2$ plane, Regions I and III are
  semi-infinite. The system is also infinite in extent along the entire $z$ axis. In (a) we illustrate the single-mode
  approximation developed in the text: the field profiles in Regions I (solid line) and II (dashed), calculated for
  $\omega R/c=800$, are well matched. In (b), the intensity is overlayed in grayscale, showing the overlap with the
  metal (dielectric constant $\epsilon_i$) and the outer dielectric ($\epsilon_o$). Arrows indicate incident and
  reflected fields in Region I, and transmitted field in Region III.
}
\end{figure}
The geometry we study is that of propagation about a rounded edge, as shown in Fig. 1. SPPs are incident along
the interface from Region I, and propagate counterclockwise through the bend in Region II, in the direction of
Region III. We calculate the energy-reflection and transmission coefficients, as well as bend-induced radiation
losses. For lossy metals absorption losses are also evaluated.

Our approach exploits known expressions for the SPP fields in each region and matches them at the two ends of
the bend. The procedure differs from related numerical techniques \cite{hafner} in that we consider the SPP
itself as the incident wave, with the goal of manipulating it as a well-defined quasi-particle in a non-trivial
geometry. Favorable conditions for this will be seen to emerge in the short-wavelength limit, and we use this to
arrive at analytic expressions.

The solutions for SPP propagation on an infinite flat surface and on cylindrical surfaces are known analytically
\cite{Sernelius}. On a flat interface, SPPs at frequency $\omega$ are two dimensional waves, decaying
exponentially into the two adjoining media, with decay constants
$\gamma_i=-{\omega}\epsilon_i\sqrt{{-1}/{(\epsilon_i+\epsilon_o)}}/{c}$ in the metal and
$\gamma_o={\omega}\epsilon_o\sqrt{{-1}/{(\epsilon_i+\epsilon_o)}}/{c}$ in the dielectric. In the limit
$\hbox{Re}[\gamma_i]R\gg1$ the interference of SPPs in Regions I and III is negligible, allowing us to use the
infinite flat surface 2D solution in these regions.

We construct the solution in Region II using the known solutions for SPPs propagating around the perimeter of an
infinitely long cylindrical metal rod of radius $R$ \cite{Sernelius}. Here, the magnetic field is given by
$${\bf B}=-i\hat{\bf z}\sum_{\{n\}} \Bigl[A_n^+e^{+in\phi}-A_n^-e^{-in\phi}\Bigr] \sqrt{\epsilon_i} J_n(k_i r
)e^{-i\omega t}$$ where $k_i={\omega}\sqrt{\epsilon_i}/{c}$ and $J_n$ is the Bessel function. The set $\{n\}$ is
determined by the metal boundary matching equation,
\begin{equation}\label{1}\frac{1}{k_i}\frac{J_n'(k_iR)}{J_n(k_iR)}
=\frac{1}{k_o}\frac{H_n^{(1)'}(k_o R)}{H_n^{(1)}(k_o R)}
\end{equation}
where $k_o={\omega}\sqrt{\epsilon_o}/{c}$, $H_n^{(1)}$ is the Hankel function of the first kind, and the prime
denotes differentiation with respect to the argument. Assuming $\omega$ real, one finds $n$ to be complex, as a
consequence of radiation loss and absorption in the bend. Since the wave depends on $\phi$ as $\exp[\pm i
n\phi]=\exp\bigl[\pm\bigl(i\hbox{Re}[n]-\hbox{Im}[n]\bigr)\phi\bigr]$, where $\phi$ is measured from the $x_1$
axis, only solutions with $\hbox{Im}[n]\geq0$ are physical for damped propagation.

Solving exactly for the transmission and reflection coefficients requires matching an infinite number of
solutions at the boundaries along the $x_1$- and $x_2$-axes separately. However, it is possible to render this
problem tractable by a few simple approximations. Noting that the incident SPP carries momentum proportional to
$k={\omega}\sqrt{\epsilon_o\epsilon_i/(\epsilon_o+\epsilon_i)}/{c}$, in the short wavelength limit ${\omega
R}/{c}\gg1$ its angular momentum with respect to the origin is approximately equal to $\hbox{Re}[k]R$. In Region
II the solution has angular momentum proportional to the various $n$-values. Conservation of angular momentum
dictates that the incident SPP couple predominantly to that cylindrical mode with $n$ closest in value to $kR$.
Therefore it is necessary to consider only a single term of the expansion. Formally, this is shown by examining
the set $\{n\}$ and noting that it contains an element $m$ which minimizes the mismatch between the field
profiles perpendicular to the surface. The role of angular momentum conservation in this matching problem is
analogous to that of tangential momentum conservation in refraction at a dielectric interface. We call the
clockwise and counterclockwise modes corresponding to $m$ the fundamental modes. In the short wavelength limit
of a fundamental mode $n=m\approx kR$, and the decay rate is $\sqrt{m^2/R^2-k_i^2}\approx\gamma_i.$ Thus, $J_m(k
_i r)\sim \exp[\gamma_i r]$ near the interface, identical to the fields in the metal in Regions I and III.
Similarly, in the dielectric $H_m ^{(1)}(k _o r)\sim \exp[-\gamma_o r]$ \cite{approx}.

The modes $n\ne m$ have decay rates not as close to $\hbox{Re}[\gamma_i]$ as that of the fundamental mode's. For
this reason, it is possible to assume that in the short wavelength limit the incident SPPs couple predominantly
to the fundamental modes and ignore other mode coupling. In order to satisfy the standard Maxwell boundary
conditions it is therefore necessary to match only a small number of solutions at a single point on each axis,
at a distance $R$ from the origin. The boundary conditions are thus also satisfied approximately over the entire
extent of the axes. As can be seen from Fig. 1, the mode mismatch at the boundaries may be very slight. The
problem has now essentially become one dimensional (1D), analogous to scattering from a 1D finite potential well
\cite{Mekis}. Since the allowed $m$-values are always complex, bound-state solutions in the well do not exist.
This distinguishes the SPP on a bent surface from waveguide bends enclosed on all sides by infinite potential
walls \cite{sols}.

Applying the appropriate boundary conditions to the fields at the two boundaries results in the familiar
expression for the transmittance
\begin{eqnarray} \textsf{T}=\Biggl|\frac{4mkR}{-e^{i m\theta}(m-k R)^2+e^{-i m
\theta}(m+k R)^2}\Biggr|^2.\end{eqnarray}
When the losses in the metal are accounted for, $\hbox{Im}[m]$
increases with $R$, such that when $\hbox{Im}[m]\theta\gg1$ the transmittance becomes
\begin{equation}\label{3}\textsf{T}\approx16\frac{|m k R|^2}{|m+k
    R|^4}e^{-2\hbox{\footnotesize Im}[m]\theta}.
\end{equation}
The reflectance $\textsf{R}$ may be obtained in a similar manner. In the limit $\omega
R/c\rightarrow\infty$ these expressions become exact.

For lossless metals, the bend-induced radiation losses are simply given by
$\textsf{A}\equiv1-\textsf{T}-\textsf{R}$. Accounting for absorption in the metal we find that $\textsf{A}$ now
includes both radiation and absorption losses. We extract the radiation losses by integrating the Poynting
vector ${\bf S}$ for unit incident flux in Region II at $r\rightarrow\infty$:
\begin{equation}\label{4}\textsf{P}\equiv\int_{\phi_0}^{\theta+{\phi_0}}{\bf
    S}\cdot\hat{\bf r}rd\phi.
\end{equation}
Since the radiation carries angular momentum with respect to the origin, the energy radiated into the far-field
from the surface at $\phi=0$ propagates at an angle $\phi_0$ with respect to the $x_1$ axis, setting the lower
integration limit in (\ref{4}). In the short-wavelength limit only the amplitude of the forward-propagating mode
is significant, therefore the radiation losses are well approximated by integrating only this mode. To obtain
$\phi_0$ we use a stationary phase approximation. The position-dependent phase is $\Phi=k_o r+\hbox{Re}[m]\phi$,
and the vector normal to a surface of constant phase is ${\bf v}({\bf r})={\bf \nabla}\Phi=k_o\hat{\bf
r}+{\hbox{Re}[m]}/{r}\hat{\bf e}_{\phi}$. The change in angle as the wave propagates a radial
distance $\delta r$ is $\delta\phi={\hbox{Re}[m]/ (k_o r^2)}\delta r$, giving $\phi_0=\int_R^{\infty}
{\hbox{Re}[m]/ (k_o r^2)}dr=\hbox{Re}[m] /(k_oR)$.

\begin{figure}[!hbt]
\centering{\includegraphics[width=10.5cm]{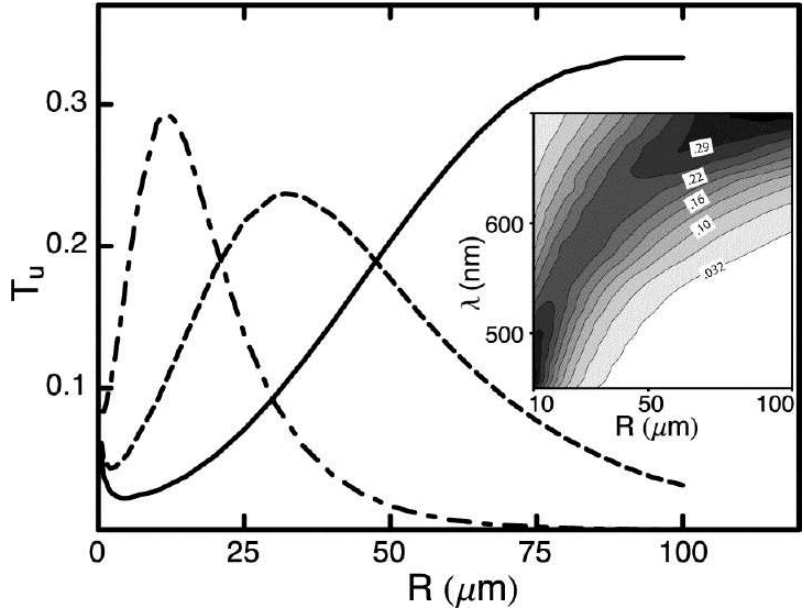}}
\caption{\label{fig2} 
The upper bound for the transmittance, $\textsf{T}_u$, plotted
  for a silver-air interface with $\theta=90^{\circ}$, as a function
  of bend radius $R$ for wavelengths of $\lambda=500$nm (dashed-dotted),
$\lambda=600$nm (dashed) and $\lambda=700$nm (solid). Inset:
$\textsf{T}_u$ in grayscale as a function of $R$ and $\lambda$.
}
\end{figure}
We have carried out calculations for typical values of silver ($\epsilon_i=-15+i0.5$) in air ($\epsilon_o=1$)
when ${\omega R}/{c}=800$ and $\theta=90^{\circ}$. Ignoring the losses in the metal we find $\textsf{T}=0.997$,
$\textsf{R}=1.19\times10^{-8}$ and $\textsf{P}\approx0.003$. When the losses are included the results change
drastically to $\textsf{T}=0.0516$, $\textsf{R}=1.18\times10^{-6}$ and $\textsf{P}\approx0.00282$, indicating
that most of the energy is lost to absorption in the metal. Comparing the latter overall absorption and
radiation losses to the energy absorbed when SPPs propagate the equivalent arc distance on a flat surface, we
find that propagation on a non-planar interface may result in $lower$ losses. We explain this counterintuitive
result using an analogous picture of semi-classical motion under an effective potential in a central potential
field. In the short wavelength and large angular momentum limit the SPP fields propagating on the curved
interface sample less of the metal volume than that available when propagating on a flat interface, hence the
reduced absorption.

We evaluate the accuracy of our result by examining the coupling efficiency $\Delta$ of a single mode on a flat
interface to a fundamental mode $m$. We define this by
\begin{equation}\label{4a}\Delta^2\equiv
\frac{{\int_{R}^{R+\eta\gamma_o^{-1}}\Biggl|\frac{H_m^{(1)}(k_o r)}{H_m^{(1)}(k_o R)}-\exp[-\gamma_o
(r-R)]\Biggr|^2dr}}{{\int_{R}^{R+\eta\gamma_o^{-1}}\big|\exp[-\gamma_o (r-R)]\bigr|^2dr}}
\end{equation}
where $\eta=O(1)$. We find that the condition $\Delta\ll 1$ constitutes a stricter criterion for the validity of
our approximation. When the latter holds, the incident SPP couples predominantly to the fundamental modes,
making the approach described above self-consistent. For example, when $\eta=3$, $\epsilon_i=-15$, and
$\epsilon_o=1$, $\Delta^{2}=0.002$ for ${\omega R}/{c}=800$, rendering our result applicable. On the other hand,
for ${\omega R}/{c}=100$ we obtain $\Delta^{2}=0.3$, signifying that the expression is not reliable because the
coupling to modes $n$ other than the fundamental can no longer be neglected. In this regime a more physical
quantity is the upper bound for the transmittance, given from (3) by
\begin{equation}\label{ub}\textsf{T}_u=\exp\bigl(-2\hbox{Im}[m]\theta\bigr).\end{equation}
Here we neglect reflections at the boundaries between the different regions, thus excluding interference with
the counter-propagating mode in Region II. To understand why this is an upper bound, recall that in the
wavelength range of interest, where the metal is not very lossy, $\hbox{Im}[n]>\hbox{Im}[m]$. Since the wave
depends on $n$ as $\exp[\pm i n\phi]$, modes with large $\hbox{Im}[n]$ decay rapidly. Thus, the transmission in
the presence of coupling to non-fundamental modes does not exceed $\textsf{T}_u$, and the latter is a true upper
bound. Fig. 2 is a plot of $\textsf{T}_u$. A peak is clearly visible, moving to higher values of $R$ as the
wavelength increases. To the right of the peak, at large radii of curvature absorption losses in the metal
dominate, and the maximum transmittance decreases with increasing radius. To the left of the peak radiation due
to the high curvature is the dominant loss mechanism, leading to a rapid drop in $\textsf{T}_u$. At very high
curvature ($R\leq10\mu$m) there is a change in trend, and $\textsf{T}_u$ starts to {\em increase} with
decreasing $R$ (see Inset.) When calculating the radiation loss per arclength, we find that for this range of
radii it increases slower than elsewhere, allowing $\textsf{T}_u$ to increase even as $R$ attains very small
values. This anomalous behavior can be observed for all wavelengths, and is independent of the dispersion in the
metal.

The formalism developed here may also be used to analyze the complementary reversed geometry, where the metal
occupies three quadrants in space, and the SPPs propagate around a dielectric void in it. Surprisingly, in this
case we find that in the single mode approximation SPP propagation around the bend is non-radiative. Separate
work will address radiation and absorption processes in this system in greater detail.

In summary, we have analyzed the scattering of SPPs at a curved metal-dielectric interface in the short
wavelength limit. Utilizing an analogy to a quantum mechanical 1D finite square well we obtained the energy
transmission and reflection coefficients. Interestingly, propagation on a curved interface may result in lower
losses than at a flat metallic surface, due to the unique field distributions which arise in our system. An
expression for an upper bound on the transmittance was also obtained, showing that at high curvature radiation
is the main loss mechanism, while at low curvature material losses dominate. An unexpected behavior where the
maximum transmittance increases with curvature was also observed. We explain this as an interplay between
various loss rates in the system. These results shed new light on the mesoscopic behavior of SPPs, and should
play an important role in the design and optimization of SPP devices. Future work will address SPP propagation
in waveguides, splitters and interferometers.

J.U.N. acknowledges support from NSF Grant ECS-02-39332; K.H. and M.D. acknowledge support from NSF Grant
DMR-02-39273 and ARO Grant DAAD19-02-1-0286.


\begin{thebibliography}{9}
\bibitem {approx} To show this we use the Bessel equation
$r^2d^2f/ dr^2+r{df/dr}+(k_{i,o}^2r^2-n^2)f=0,$ to which $J_m(k _i r)$ and $H_m ^{(1)}(k _o r)$ are solutions,
respectively. With $r=R+x$, in the limit $x\ll R$ this reduces to $R^2{d^2f/ dx^2}+(k_{i,o}^2 R^2-n^2)f=0$, to
which the solutions are exponentials.
\end{thebibliography}
\end{document}